\documentclass[prd,twocolumn,nopacs,floatfix,amsmath,nofootinbib,amssymb,preprintnumbers]{revtex4-2}
\usepackage{graphicx,color,dcolumn,booktabs,diagbox}
\usepackage{txfonts}
\usepackage{fontspec}
\usepackage{cancel}
\usepackage{slashed}
\usepackage{epstopdf}
\usepackage{multirow}
\usepackage{tikz}
\usetikzlibrary{decorations.pathmorphing}
\usepackage{bm}
\usepackage{braket}
\usepackage{pifont}
\usepackage{adjustbox}
\usepackage{rotating}
\usepackage{verbatim}
\usepackage{appendix}
\usepackage{graphicx,xcolor,dcolumn,booktabs}
\usepackage[colorlinks,
            citecolor=blue,
            anchorcolor=red,
            menucolor=red,
            linkcolor=red,
            filecolor=red,
            runcolor=red,
            urlcolor=blue,
            frenchlinks=red]{hyperref}

\usepackage{placeins}
\AddToHook{env/thebibliography/begin}{\interlinepenalty=10000}
\usepackage{newunicodechar}
\newunicodechar{π}{\ensuremath{\pi}}
\newunicodechar{η}{\ensuremath{\eta}}
\newunicodechar{Λ}{\ensuremath{\Lambda}}
\newunicodechar{−}{-}
\makeatletter
\def\@bibstyle{x3790-apsrev4-2}
\makeatother
\newif\ifincludeangular
\includeangulartrue

\graphicspath{{Figures/}}
\makeatletter
\@ifundefined{saved@reqcolroom}{%
  \newdimen\saved@reqcolroom
  \def\@flcheckspace#1#2{%
    \@reqcolroom\saved@reqcolroom
    \advance\@reqcolroom
      \ifx#2\@empty\textfloatsep\else\floatsep\fi
    \ifdim\@colroom>\@reqcolroom
      \ifdim#1>\ht\@currbox
        \@tempswatrue
      \else
        \ifnum\@fpstype<\sixt@@n\@tempswatrue\fi
      \fi
    \fi}
}{}
\makeatother
\begin{document}
\title{BESIII $X(3790)$ enhancement as a reflection of higher charmed mesons}
\author{Peng-Yu Sun$^{1,2,4}$}\email{sunpy2025@lzu.edu.cn}
\author{Tian-Le Gao$^{1,2,4}$}\email{gaotl2025@lzu.edu.cn}
\author{Xiang Liu$^{1,2,3,4}$}
\email{xiangliu@lzu.edu.cn}
\iffalse
\affiliation{
$^1$School of Physical Science and Technology, Lanzhou University, Lanzhou 730000, China\\
$^2$Lanzhou Center for Theoretical Physics, Lanzhou University, Lanzhou 730000, China\\
$^3$Key Laboratory of Quantum Theory and Applications of MoE, Lanzhou University,
Lanzhou 730000, China\\
$^4$Key Laboratory of Theoretical Physics of Gansu Province, Lanzhou University, Lanzhou 730000, China\\
$^5$MoE Frontiers Science Center for Rare Isotopes, Lanzhou University, Lanzhou 730000, China\\
$^6$Research Center for Hadron and CSR Physics, Lanzhou University and Institute of Modern Physics of CAS, Lanzhou 730000, China}
\fi
\affiliation{
$^1$School of Physical Science and Technology, Lanzhou University, Lanzhou 730000, China\\
$^2$Lanzhou Center for Theoretical Physics, Key Laboratory of Theoretical Physics of Gansu Province, Key Laboratory of Quantum Theory and Applications of MoE, Gansu Provincial Research Center for Basic Disciplines of Quantum Physics, Lanzhou University, Lanzhou 730000, China\\
$^3$MoE Frontiers Science Center for Rare Isotopes, Lanzhou University, Lanzhou 730000, China\\
$^4$Research Center for Hadron and CSR Physics, Lanzhou University and Institute of Modern Physics of CAS, Lanzhou 730000, China}

\date{October 15, 2026}

\begin{abstract}

In this work, we show that the $X(3790)$ reported by the BESIII Collaboration is unlikely to be a genuine resonance, a conclusion that follows from the incompleteness of the BESIII experimental analysis. Although BESIII included the established charmed mesons that couple to $D\pi$, as listed by the Particle Data Group, and incorporated their reflections in the $D^0D^-$ invariant mass spectrum, this fit to the experimental data does not exhaust the charmed mesons that can contribute to the $D^0D^-$ invariant mass spectrum through reflections, as can be explicitly seen from a theoretical charmed-meson spectroscopy. We find that a predicted charmed meson, either $D(3\,{}^3S_1)$ or $D(2\,{}^3D_1)$, can play a crucial role in mimicking the $X(3790)$ structure through the corresponding reflection. This warrants caution in interpreting the $X(3790)$ as a genuine resonance. Moreover, such an interpretation is not supported by most theoretical studies of $D\bar{D}$ hadronic molecules and compact $c\bar{c}d\bar{u}$ tetraquarks currently available.

\end{abstract}

\maketitle

% Compact display spacing; preserve all font sizes and formula expressions.
\setlength{\abovedisplayskip}{3pt plus 1pt minus 1pt}
\setlength{\belowdisplayskip}{3pt plus 1pt minus 1pt}
\setlength{\abovedisplayshortskip}{2pt plus 1pt}
\setlength{\belowdisplayshortskip}{3pt plus 1pt minus 1pt}

\flushbottom\clubpenalty=2000\widowpenalty=2000\brokenpenalty=10000
\section{Introduction}
\label{sec:intro}

Electron–positron annihilation into open-charm final states is an ideal platform to probe hadron spectroscopy \cite{Wang:2025dur,Bai:2026atm}.
Very recently, a charged charmoniumlike enhancement, denoted as $X(3790)$, was reported by the BESIII Collaboration in $e^+e^-\to D^0 D^-\pi^+$, based on an analysis of a data sample of $1667\,\mathrm{pb}^{-1}$ collected at the collision energy $W=4.682\,\mathrm{GeV}$~\cite{BESIII:2026emz}. Since its mass measured by a Flatt\'e parameterization~\cite{Flatte:1976xu} is $3794.7\pm6.0\pm 8.7\,\mathrm{MeV}$ and its quantum numbers are determined to be $J^P=1^-$, it is natural to conjecture whether this enhancement could be an exotic hidden-charm hadronic state. In fact, such an assignment is not supported by current theoretical studies of $D\bar{D}$ molecular states and compact $c\bar{c}d\bar{u}$ tetraquarks \cite{Lu:2025zae,Ebert:2008kb,Wang:2024gvf,Wang:2021qus}. The reported $X(3790)$ is hard to assign to exotic hadronic states like a molecular or tetraquark state.

Faced with this situation, we should reexamine the details of the analysis performed by BESIII. In their analysis, they considered several excited charmed meson contributions (e.g., $D_1^*(2600)$, $D_1^*(2760)$, and $D_3^*(2750)$) in the $D\pi$ invariant mass spectrum and evaluated their kinematic reflections into the $D^0 D^-$ invariant mass spectrum~\cite{BESIII:2026emz}. BESIII found that the enhancement near $3790\,\mathrm{MeV}$ in the $D^0 D^-$ invariant mass spectrum cannot be described by these reflections alone. Thus, they introduced an additional $X(3790)$ contribution in their fit. This is the logic behind the announcement of this new observation.

The experimental charmed meson spectrum remains incomplete \cite{Chen:2016spr}, so a reflection test should also examine the higher excitations predicted by spectroscopy~\cite{ParticleDataGroup:2024cfk,Song:2015fha}.

In fact, the Lanzhou group noticed this effect in their studies of the reported charmoniumlike structures $Z_c(3885)$ and $Z_c(4025)$. They used reflections to describe these enhancements, with the off-shell $D_1(2420)$ contributing to $D\bar{D}^*\pi$ and $D^*\bar{D}^*\pi$, including interference with a nonpeaking amplitude \cite{Wang:2020axi}. Moreover, the dependence of these enhancements on collision energy follows from the position of the resonance band relative to the Dalitz plot boundary \cite{Wang:2020dmv}. The same approach was applied to $Z_{cs}(3985)$ via the chain $Y(4660) \to D_s^{*-} D_{s2}^*(2573)^+ \to D_s^{*-} D^0 K^+$, where a combined description of the kaon recoil spectra at five energies gives a pronounced enhancement near the $D_s^* D_{s2}^*(2573)$ threshold \cite{Wang:2020kej}.

\begin{figure*}[!t]
\centering
\includegraphics[width=0.95\textwidth]{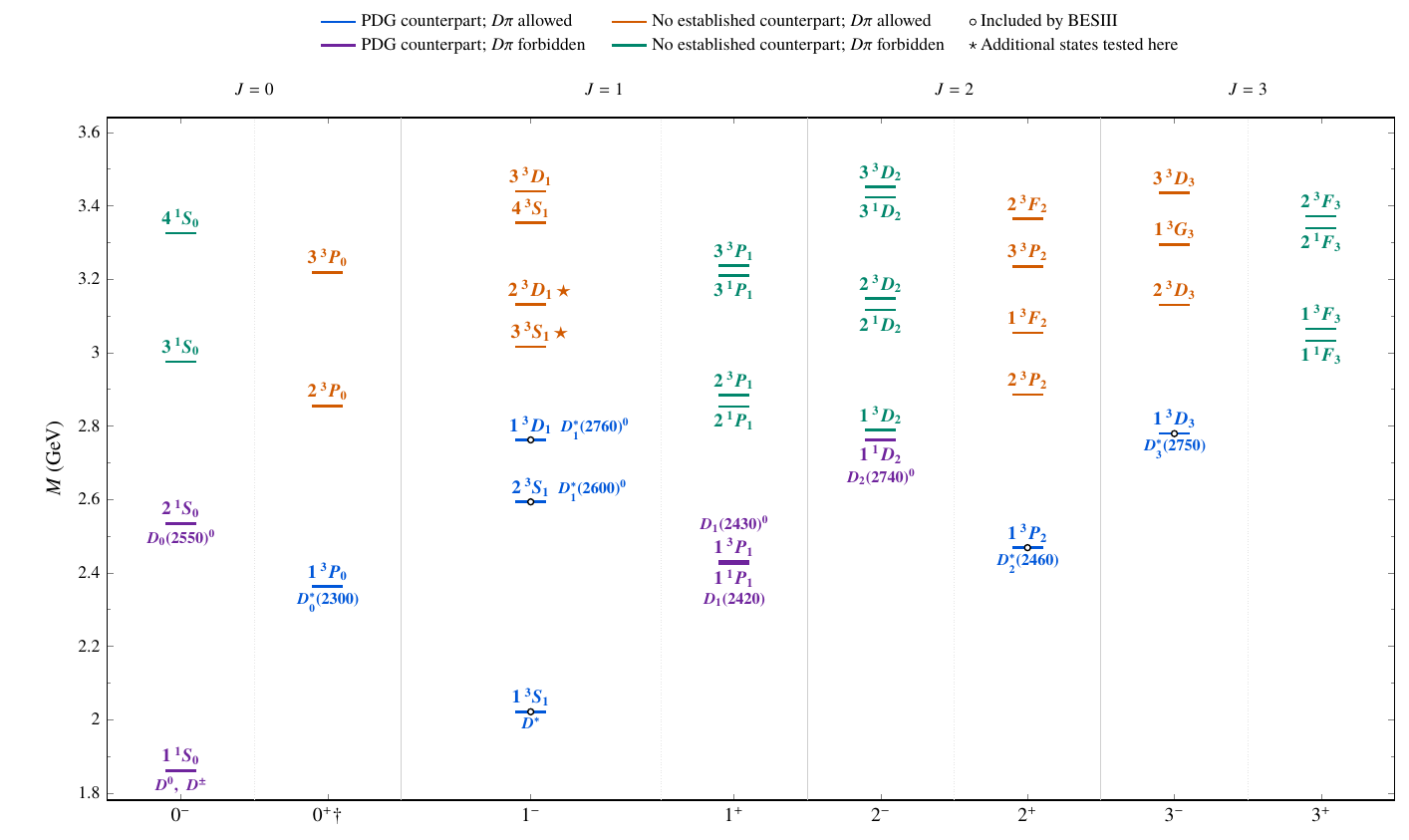}
\caption{Theoretical charmed meson mass spectrum for $J\leq3$. Blue and orange lines denote states that can decay to $D\pi$, with and without adopted PDG counterparts, respectively; purple and green lines denote the corresponding states for which this decay is forbidden~\cite{ParticleDataGroup:2024cfk}. Circles mark the five families included by BESIII~\cite{BESIII:2026emz}, and stars indicate the two additional candidates considered here (Table~\ref{tab:masswidth}). The spectroscopic labels give the dominant $n\,{}^{2S+1}L_J$ component. $\dagger$: a $0^+$ state can decay to $D\pi$ but cannot be produced with a pseudoscalar $D$ from a $1^-$ source.}
\label{fig:spectrum}
\end{figure*}

In this work, we propose that the reported $X(3790)$ enhancement can be due to an incomplete experimental analysis. Therefore, the $X(3790)$ is unlikely to be a genuine resonance. To illustrate this point, we first present the whole spectroscopy of charmed mesons obtained from a potential model and compare it with the charmed mesons collected by the Particle Data Group (PDG)~\cite{ParticleDataGroup:2024cfk}. This comparison shows that the charmed mesons adopted in the BESIII analysis are only a subset of the predicted charmed mesons allowed to couple to $D\pi$. When the $3\,{}^3S_1$ and $2\,{}^3D_1$ charmed-meson states (in the $n\,{}^{2S+1}L_J$ notation, both with $J^P=1^-$), which are not included in the BESIII analysis, are taken into account, we find that their reflections can play a crucial role in mimicking the $X(3790)$ enhancement in the $D^0D^-$ invariant mass spectrum. Detailed analyses are presented in the following sections.

This paper is organized as follows. After the introduction, we present the complete charmed meson spectrum and illustrate how to select the allowed charmed mesons that contribute to the $D^0D^-$ invariant mass spectrum through reflections (see Sec.~\ref{sec:spectrum}). In Sec.~\ref{sec:framework}, we introduce the analysis framework and present the final fit results for the three invariant mass spectra of $e^+e^-\to D^0D^-\pi^+$. This study supports the scenario that the $X(3790)$ is not a genuine resonance. Finally, we end with a short summary.

\section{Charmed meson spectroscopy and selection of charmed mesons for the analysis}
\label{sec:spectrum}

\begin{table}[!b]
\caption{Theoretical masses $M$ and summed two-body widths $\Gamma_{\rm sel}$ (MeV), and inclusive $D\pi$ branching fractions $B_{D\pi}$ (\%). Superscripts $\pm$ and $0$ denote charged and neutral parents, respectively; charge-conjugate states have equal values. Stars mark the additional states used in the fits. Calculation details are given in Appendix~\ref{app:modelinputs}.}
\label{tab:masswidth}
\renewcommand{\arraystretch}{1.05}
\centering
\setlength{\tabcolsep}{0pt}
\begin{tabular*}{\linewidth}{@{\extracolsep{\fill}}lD{.}{.}{4.2}*{4}{D{.}{.}{3.2}}@{}}
\toprule
State & \multicolumn{1}{c}{$M$} & \multicolumn{1}{c}{$\Gamma_{\rm sel}^{\pm}$} & \multicolumn{1}{c}{$\Gamma_{\rm sel}^{0}$} & \multicolumn{1}{c}{$B_{D\pi}^{\pm}$} & \multicolumn{1}{c}{$B_{D\pi}^{0}$}\\
\midrule
$1\,{}^3S_1$ & 2021.02 & 0.09 & 0.09 & 100.00 & 100.00 \\
$1\,{}^3P_2$ & 2468.18 & 11.81 & 11.81 & 69.32 & 69.32 \\
$1\,{}^3D_3$ & 2778.95 & 17.03 & 17.03 & 45.03 & 45.01 \\
$1\,{}^3D_1$ & 2761.79 & 367.80 & 367.96 & 21.48 & 21.47 \\
$2\,{}^3S_1$ & 2592.95 & 43.10 & 43.13 & 21.37 & 21.36 \\
$3\,{}^3S_1\star$ & 3014.82 & 89.00 & 89.08 & 12.64 & 12.63 \\
$2\,{}^3D_1\star$ & 3130.59 & 325.43 & 325.48 & 11.50 & 11.50 \\
$2\,{}^3D_3$ & 3129.50 & 44.34 & 44.35 & 7.23 & 7.22 \\
$1\,{}^3F_2$ & 3053.48 & 401.40 & 401.78 & 5.86 & 5.86 \\
\bottomrule
\end{tabular*}
\end{table}

To understand whether reflections can account for $X(3790)$, we first need a sufficiently complete picture of the charmed mesons that may contribute to $e^+e^-\to D^0D^-\pi^+$. Their masses and decay properties determine which $D\pi$ contributions should be considered and how they may affect the $D^0D^-$ invariant mass spectrum.
 This motivates a study of the excited charmed meson spectrum beyond the established states already included in the experimental analysis.

We therefore calculate radial and orbital charmed meson excitations with the Modified Godfrey--Isgur (MGI) model~\cite{Godfrey:1985xj,Song:2015fha}. As shown in Fig.\ref{fig:spectrum}, the model gives a good overall description of the masses of established charmed mesons and predicts additional states that have not yet been established experimentally~\cite{ParticleDataGroup:2024cfk,Song:2015fha}. We then calculate their two-body strong decays using the MGI radial wave functions and the Quark Pair Creation (QPC) model, restricting the channels to those allowed by the Okubo--Zweig--Iizuka (OZI) rule~\cite{LeYaouanc:1972vsx,LeYaouanc:1973ldf,Song:2015fha,Song:2015nia}. The model inputs and conventions are given in Appendix~\ref{app:modelinputs}. Together, these calculations allow us to assess which additional states may contribute to the measured $D\pi$ invariant mass spectrum.

Fig. \ref{fig:spectrum} shows that experimental information on the charmed meson spectrum remains sparse above about $2.8$~GeV. BESIII has included the established $D^*$, $D_2^*(2460)$, $D_1^*(2600)$, $D_1^*(2760)$, and $D_3^*(2750)$ families and their charge partners~\cite{BESIII:2026emz}, but further states predicted by the model can also couple to $D\pi$. Their possible reflections should therefore be examined. At $W=4.682$~GeV, the largest accessible $D\pi$ mass is $W-m_D\simeq2.81$--$2.82$~GeV, where $m_D$ is the spectator mass. The predicted excitations around $3.0$--$3.2$~GeV lie above these endpoints, so any contribution from them enters through off-shell tails. These tails can interfere with the lower-state and nonresonant amplitudes and thereby affect the $D^0D^-$ distribution. Whether they can account for the near-threshold enhancement must be tested with the full coherent amplitude.

We next consider which of these states are most promising, taking into account their quantum numbers, $D\pi$ branching fractions, and production partial waves. A state that decays to $D\pi$ and is produced with a pseudoscalar $D$ from a $1^-$ source must have natural parity and $J\geq1$~\cite{Jacob:1959at,Chung:1997jn}. Table~\ref{tab:masswidth} lists the masses and widths of the leading allowed candidates in order of their inclusive $D\pi$ branching fractions. The five families included by BESIII occupy the first five places for both charges, supporting the experimental choice. Among the additional states listed, $3\,{}^3S_1$ and $2\,{}^3D_1$ have the largest $D\pi$ branching fractions, approximately 12.64\% and 11.50\%, respectively. Their $J^P=1^-$ also allows them to be produced with the spectator $D$ in a relative $P$-wave. In comparison, $2\,{}^3D_3$ and $1\,{}^3F_2$ have smaller $D\pi$ branching fractions and require relative $F$-wave and $D$-wave production, respectively, with higher centrifugal barriers. These considerations favor $3\,{}^3S_1$ and $2\,{}^3D_1$ as the first additional states to investigate.

The preceding spectroscopy and decay analysis, together with the absence of an $I=1$, $J^P=1^-$ charged hidden-charm candidate compatible with $X(3790)$ in the molecular and compact tetraquark calculations considered here~\cite{Lu:2025zae,Ebert:2008kb,Wang:2024gvf,Wang:2021qus}, motivates a further test of the reflection mechanism. Table~\ref{tab:schemes} summarizes the three fit schemes, including nonresonant (NR) production.\footnote{The two additional states are chosen to illustrate the reflection framework and the implications of the incomplete charmed meson spectrum used in the experimental fit. Other allowed states may also contribute, so a more comprehensive experimental amplitude analysis is needed.} In the next section, we construct the corresponding amplitudes and fit the $D^0D^-$, $D^0\pi^+$, and $D^-\pi^+$ spectra simultaneously. This allows us to test whether the additional contributions can explain $X(3790)$ through reflections without introducing an additional resonance in the $D^0D^-$ channel.

\begin{table}[!htb]
\caption{Fit schemes. Common contributions are identical in all three schemes. Charmed meson masses and widths are from Table~\ref{tab:masswidth}; $m_X$, $g_X$, and $J^P$ are from BESIII~\cite{BESIII:2026emz}, while $c_X$ is fitted.}
\label{tab:schemes}
\centering\small
\renewcommand{\arraystretch}{1.05}
\setlength{\tabcolsep}{0pt}
\begin{tabular*}{\linewidth}{@{\extracolsep{\fill}}lcc@{}}
\toprule
Scheme & Common contributions & Added state\\
\midrule
Scheme I & \multirow{3}{*}{\shortstack{$1\,{}^3S_1,\ 1\,{}^3P_2,\ 2\,{}^3S_1,$\\$1\,{}^3D_1,\ 1\,{}^3D_3$, NR}} & $3\,{}^3S_1$\\
Scheme II & & $2\,{}^3D_1$\\
Scheme III & & $X(3790)$\\
\bottomrule
\end{tabular*}
\end{table}

\section{Reflection mechanism and phenomenological analysis}
\label{sec:framework}
\label{sec:results}

\subsection{The framework of analysis}
\label{subsec:fitframework}

\begin{figure*}[!t]
\centering
\includegraphics[width=\textwidth]{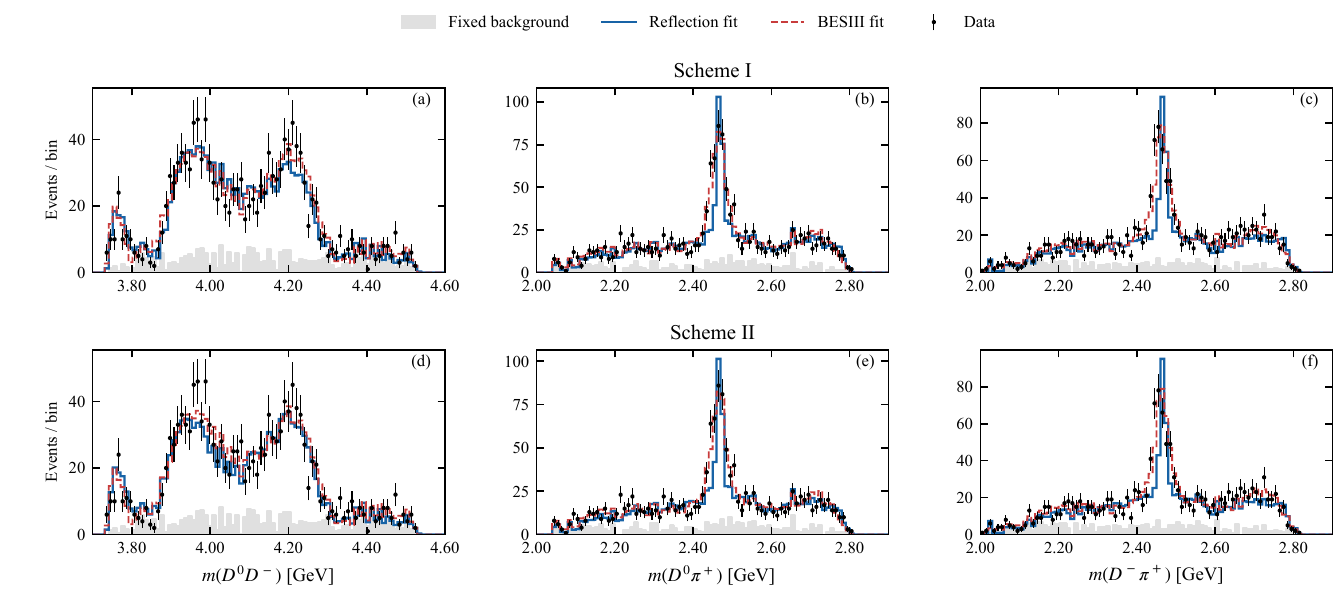}
\caption{Simultaneous fits to the $D^0D^-$, $D^0\pi^+$, and $D^-\pi^+$ spectra (columns) in Scheme I (top row) and Scheme II (bottom row), defined in Table~\ref{tab:schemes}. Blue curves show our fits including background; black points are the data. Red dashed curves show the BESIII fit with $X(3790)$, and gray areas indicate the experimental background~\cite{BESIII:2026emz}.}
\label{fig:s3}
\end{figure*}

For $D^0(p_1)D^-(p_2)\pi^+(p_3)$, let $s=(p_1+p_2)^2$, $t=(p_1+p_3)^2$, and $u=(p_2+p_3)^2$, with $s+t+u=W^2+\sum_i m_i^2$, where $m_i$ are the final-state masses. For the selected $R\to D\pi$ states, the $D^-R^+$ and $D^0\bar R^0$ chains interfere coherently. Angular momentum and parity conservation select natural-parity $J\geq1$ states with $L_{\rm prod}=\ell_{\rm dec}=J$~\cite{Jacob:1959at,Chung:1997jn}. We write the amplitude for a vector source as
\begin{equation}
\begin{aligned}
 \mathcal A_M&=\bm\epsilon_M\cdot(\bm p_1\times\bm p_2)\,G(s,t,u),\\
 G(s,t,u)&=c_{\rm NR}+\sum_R c_R
 \bigl[f_{R,+}(s,t)+f_{R,\bar0}(s,u)\bigr],
\end{aligned}
\label{eq:amplitude}
\end{equation}
\noindent where $\bm\epsilon_M$ is the source polarization and momenta are in the overall center-of-mass frame. The vector product is the center-of-mass reduction of the parity-conserving Levi-Civita structure~\cite{JPAC:2020umo}. The fitted coefficients contain production strengths and strong phases; $c_{\rm NR}$ is constant. Sharing $c_R$ between charge channels implements BESIII's isospin constraint~\cite{BESIII:2026emz}. The plus sign follows our chain phase convention with a common normal vector.

Factoring out this common vector product gives our reduced spin-$J$ amplitude,
\begin{equation}
 f_{R,c}(s,x_c)=
 \frac{(p_cq_c)^{J-1}P'_J(z_c)\mathcal B_J(p_c)\mathcal B_J(q_c)}
 {x_c-M_{R,c}^2+iM_{R,c}\Gamma_{R,c}(\sqrt{x_c})},
 \label{eq:chain}
\end{equation}
\noindent where $x_+=t$ and $x_{\bar0}=u$. The momenta $p_c$ and $q_c$ describe production in the overall center-of-mass frame and decay in the resonance frame, respectively; $z_c$ is the daughter--spectator angle cosine in the latter, and $P'_J$ is a Legendre polynomial derivative. The barrier forms and radius $r_B$ follow Refs.~\cite{VonHippel:1972fg,LHCb:2016lxy,BESIII:2019dme}. We use $\mathcal B_J(0)=1$, absorbing the constants from both vertices, angular normalization and the $J$-dependent helicity sign into $c_R$~\cite{Jacob:1959at,Chung:1997jn}. QPC partial widths determine $\Gamma_{R,c}(m)$ through Eq.~\eqref{eq:widthenergy}.

Three-body phase space projection gives the spectra; for $D^0D^-$~\cite{ParticleDataGroup:2024cfk},
\begin{equation}
 \frac{d\sigma}{dm(D^0D^-)}\propto
 2m(D^0D^-)\int_{t_{\min}(s)}^{t_{\max}(s)}dt\,
 |\bm p_1\times\bm p_2|^2|G(s,t,u)|^2,
 \label{eq:reflectionprojection}
\end{equation}
\noindent with analogous expressions for $D\pi$. The physical $t$ and $u$ intervals at each $s$ allow off-shell $D\pi$ contributions to produce a reflected enhancement through interference. All projections include $m(D^0\pi^+)>2.04$~GeV.

In Scheme III, Eq.~\eqref{eq:amplitude} includes $c_Xf_X(s)$, with~\cite{Flatte:1976xu,BESIII:2026emz}
\begin{equation}
 f_X(s)=\frac{\mathcal B_1(p_X)\mathcal B_1(q_X)}
 {s-m_X^2+i g_Xm_Xq_X/\sqrt{s}},
 \label{eq:xbenchmark}
\end{equation}
\noindent where $p_X$ and $q_X$ are the production and decay momenta in the overall center-of-mass and $D^0D^-$ frames, respectively, and $\mathcal B_1(k)=[1+(kr_B)^2]^{-1/2}$. We adopt BESIII's Flatt\'e form rather than the charmed meson Breit--Wigner denominator~\cite{BESIII:2026emz}.

All fits use the $N_{\rm pt}=238$ points shown in the published spectra and fixed backgrounds. Shared complex coefficients and three independent spectrum normalizations leave 15 real parameters after fixing the overall amplitude scale and phase.

We compare the fits using the original experimental errors:
\begin{equation}
 \chi^2_{\mathcal I}=\sum_{i\in\mathcal I}
 \left(\frac{N_i^{\rm pred}-N_i^{\rm obs}}{\sigma_i^{\rm exp}}\right)^2,
 \label{eq:ddchi2}
\end{equation}
\noindent where $\sigma_i^{\rm exp}$ averages the reported upper and lower errors, and $N_i^{\rm pred}$ includes background. The sets $\mathcal I_{\rm full}$ (all 79 $D^0D^-$ points) and $\mathcal I_{\rm loc}$ (13 points in $3.73\leq m(D^0D^-)\leq3.86$~GeV) define $\chi^2_{\rm full}$ and $\chi^2_{\rm loc}$. Both are evaluated from the simultaneous fits, without refitting. The total $\chi^2_{\rm tot}=\sum_{a=1}^{3}\chi^2_a$ sums the three spectra, with $\chi^2_1=\chi^2_{\rm full}$.

\subsection{The fitting results}
\label{subsec:fitresults}

Fig. \ref{fig:s3} shows that Schemes I and II reproduce the near-threshold enhancement while also describing the two $D\pi$ spectra. Table~\ref{tab:fits} compares these results with Scheme III. The fitted coefficients are given in Appendix~\ref{app:fitcoefficients}.

\begin{table}[!htb]
\caption{Comparison of $\chi^2_{\rm full}$, $\chi^2_{\rm tot}$, and $\chi^2_{\rm loc}$ for the full $D^0D^-$ spectrum (79 points), all three spectra (238 points), and the near-threshold interval shown in Fig. \ref{fig:zoom} (13 points), respectively. All values use the original experimental errors. All three schemes include the five lower states and NR.}
\label{tab:fits}
\renewcommand{\arraystretch}{1.05}
\centering\small
\setlength{\tabcolsep}{0pt}
\begin{tabular*}{\linewidth}{@{\extracolsep{\fill}}l*{3}{D{.}{.}{3.2}}@{}}
\toprule
Scheme & \multicolumn{1}{c}{$\chi^2_{\rm full}$} & \multicolumn{1}{c}{$\chi^2_{\rm tot}$} & \multicolumn{1}{c}{$\chi^2_{\rm loc}$}\\
\midrule
Scheme I & 140.64 & 478.70 & 38.04 \\
Scheme II & 150.75 & 483.05 & 56.21 \\
Scheme III & 139.96 & 463.65 & 45.01 \\
\bottomrule
\end{tabular*}
\end{table}

\begin{figure}[!b]
\centering
\includegraphics[width=0.92\linewidth]{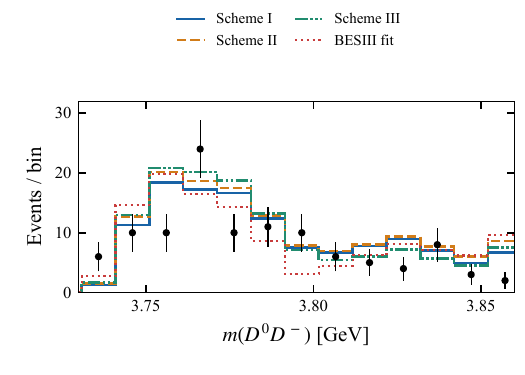}
\caption{Comparison of the fits in the near-threshold $D^0D^-$ region, including background. Blue, orange, and green curves show Schemes I, II, and III, respectively, and the red dotted curve is the BESIII fit~\cite{BESIII:2026emz}. Black points are the data with the reported errors.}
\label{fig:zoom}
\end{figure}

The comparison in Table~\ref{tab:fits} shows that Schemes I and II give $\chi^2_{\rm full}$ values close to that of Scheme III for the full $D^0D^-$ spectrum. Their $\chi^2_{\rm tot}$ values are somewhat larger when all three spectra are considered. All three fits have the same number of free parameters. The comparison indicates that the reflection mechanism can reproduce the enhancement while also describing the accompanying $D\pi$ spectra. Fig.\ref{fig:zoom} shows the near-threshold region in greater detail. Scheme I gives a smaller $\chi^2_{\rm loc}$ than Scheme III, while Scheme II gives a somewhat larger value. In particular, the agreement in the near-threshold region obtained in Scheme I supports the possibility that coherent $D\pi$ reflections can account for the reported line shape without an additional $D^0D^-$ resonance.

These results support a reflection interpretation of $X(3790)$ based on the more complete theoretical charmed meson spectrum. Within this description, the off-shell $D\pi$ contributions and their interference account for the enhancement without introducing an additional resonance in the $D^0D^-$ channel.
\ifincludeangular
The comparable descriptions of the mass spectra motivate a comparison of angular distributions. We use the three fitted amplitudes to identify angular features that could distinguish the reflection and resonance interpretations in future measurements.
\fi

% BEGIN X3790_ANGULAR: THREE_MODELS
\ifincludeangular

We label the $D^0D^-$, $D^0\pi^+$, and $D^-\pi^+$ rest frames by $a=1,2,3$, respectively, and define $\theta_1\equiv\theta_{\pi^+D^0}$, $\theta_2\equiv\theta_{D^0D^-}$, and $\theta_3\equiv\theta_{D^-D^0}$. Here $\theta_{AB}$ is the angle between particles $A$ and $B$ in the corresponding frame. With the coefficients determined by the fits to the mass spectra, the angular distributions are~\cite{ParticleDataGroup:2024cfk,Nam:2026dmr,Kim:2018hlp}
\begin{equation}
 F_a(z)=\frac{\displaystyle\int_{\mathcal D}d\Phi_3\,
 \overline{|\mathcal A_M|^2}\,\delta(z-z_a)}
 {\displaystyle\int_{\mathcal D}d\Phi_3\,
 \overline{|\mathcal A_M|^2}},
 \label{eq:angularprojection}
\end{equation}
\noindent where $z_a=\cos\theta_a$, and the bar denotes the average over initial spin states. The integration covers the full accessible phase space $\mathcal D$ subject to $m(D^0\pi^+)>2.04$~GeV. Each distribution $F_a(z)=N_{\rm sig}^{-1}dN_{\rm sig}/dz$ is normalized to the total coherent signal, so that $\int_{-1}^{1}F_a(z)dz=1$. The calculation includes the five lower states, NR, the additional contribution, and all interference terms, but not the experimental background or detector acceptance.

The three schemes give similar angular distributions in Fig. \ref{fig:angularfull}, with modest differences in the central region. In the $D^0D^-$ frame, the distributions are approximately symmetric: Scheme I has the deepest central dip, Scheme II is flatter, and the distribution in Scheme III lies between them near $\cos\theta_{\pi^+D^0}=0$. Differences also appear in the $D\pi$ frames. Resolving these features will require larger event samples and careful control of detector acceptance. Such measurements could help distinguish the reflection and resonance interpretations and complement the comparison of the mass spectra.
\fi
% END X3790_ANGULAR: THREE_MODELS

\ifincludeangular
\begin{figure}[!htb]
\centering
\includegraphics[width=0.98\linewidth]{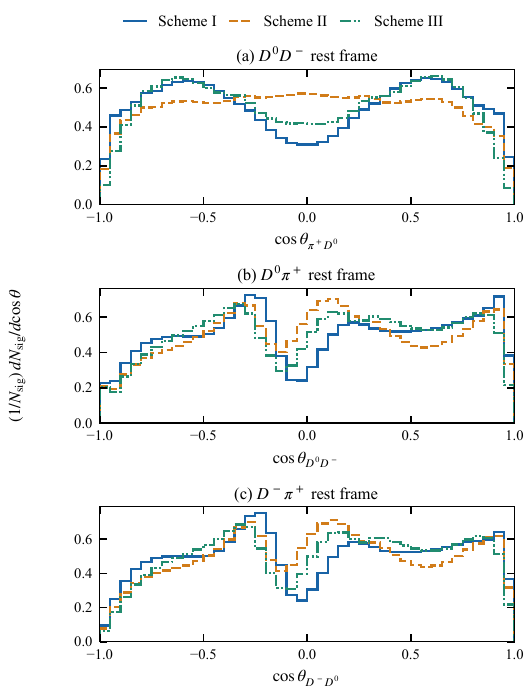}
\caption{Normalized angular distributions obtained from the full coherent amplitudes of Schemes I, II, and III, with $m(D^0\pi^+)>2.04$~GeV. The panels correspond to the $D^0D^-$, $D^0\pi^+$, and $D^-\pi^+$ rest frames. Colors follow Fig. \ref{fig:zoom}. Experimental background and detector acceptance are not included.}
\label{fig:angularfull}
\end{figure}
\fi

Measurements at different center-of-mass energies offer a complementary test, as suggested by earlier studies of $Z_c$ reflections~\cite{Wang:2020axi,Wang:2020dmv}. A change in collision energy shifts the Dalitz plot boundaries and changes the kinematic region contributing to the projection. The reflected line shape can therefore change with energy. An energy scan may help clarify the origin of $X(3790)$, while measurements in different $D\pi$ mass intervals could further test the role of the crossed-channel contributions.

\FloatBarrier

\section{Summary}
\label{sec:summary}

BESIII reported a charged $X(3790)$ enhancement in the $D^0D^-$ spectrum of $e^+e^-\to D^0D^-\pi^+$~\cite{BESIII:2026emz}, but the molecular and compact tetraquark studies discussed above do not predict a compatible $I=1$, $J^P=1^-$ candidate~\cite{Lu:2025zae,Ebert:2008kb,Wang:2024gvf,Wang:2021qus}. The experimental fit introduced an $X(3790)$ resonance only after the reflections from the PDG-listed charmed mesons included in the fit failed to describe the enhancement~\cite{BESIII:2026emz}. This restricted set of charmed mesons omits higher excitations that can couple to $D\pi$~\cite{ParticleDataGroup:2024cfk,Song:2015fha}. Together with earlier $Z_c$ and $Z_{cs}$ reflection studies~\cite{Wang:2020axi,Wang:2020dmv,Wang:2020kej}, these considerations motivate us to test whether a more complete charmed meson spectrum can explain the enhancement without an additional $D^0D^-$ resonance.

To this end, we first constructed the charmed meson spectrum within a potential model and compared it with the states collected by the Particle Data Group~\cite{ParticleDataGroup:2024cfk}. This comparison shows that the charmed mesons adopted in the BESIII analysis constitute only a subset of the predicted charmed mesons allowed to couple to $D\pi$. In particular, the $3\,{}^3S_1$ and $2\,{}^3D_1$ states, both carrying $J^P=1^-$ and both absent from the BESIII fit, can contribute to the $D^0D^-$ invariant mass spectrum through their reflections.

Including these two states in the analysis framework, we performed a combined fit to the three invariant mass spectra of $e^+e^-\to D^0D^-\pi^+$ measured by BESIII~\cite{BESIII:2026emz}. The fit shows that the reflections of the $3\,{}^3S_1$ and $2\,{}^3D_1$ charmed mesons can play a crucial role in mimicking the $X(3790)$ enhancement in the $D^0D^-$ invariant mass spectrum, without invoking an additional resonance. The resulting description of the data is consistent with the measured spectra, and the detailed fit results are presented in Sec.~\ref{sec:framework}.

These results warrant caution in interpreting the $X(3790)$ as a genuine resonance. The enhancement can be attributed to an incomplete experimental analysis, in which the restricted set of charmed mesons leaves out higher excitations that couple to $D\pi$ and whose reflections populate the $D^0D^-$ invariant mass spectrum. This conclusion is further supported by the fact that most theoretical studies of $D\bar{D}$ hadronic molecules and compact $c\bar{c}d\bar{u}$ tetraquarks do not predict a compatible candidate for the $X(3790)$~\cite{Lu:2025zae,Ebert:2008kb,Wang:2024gvf,Wang:2021qus}. A more complete charmed meson spectrum, as predicted by spectroscopy, should therefore be taken into account in future experimental analyses of this and similar enhancement structures.

\FloatBarrier

\begin{acknowledgments}
This work is supported by the Natural Science Foundation of Gansu Province (Nos. 26RCKA012 and 25JRRA799), the National Natural Science Foundation of China under Grant Nos. 12335001 and 12247101, the ``111 Center'' under Grant No. B20063, the Fundamental Research Funds for the Central Universities (lzujbky-2023-stlt01), and Lanzhou City High-Level Talent Funding. T.-L. G. is supported by the Gansu Province Postgraduate Innovation Star Program under Grant No. 2026CXZX-042. The authors used ChatGPT for language editing and for assistance with the numerical code, reviewed the output, and take responsibility for the content.

\end{acknowledgments}

\FloatBarrier

\appendix
% Keep the expanded bibliography within the compact layout at unchanged fonts.
\setlength{\abovedisplayskip}{3pt plus 1pt minus 1pt}
\setlength{\belowdisplayskip}{3pt plus 1pt minus 1pt}
\setlength{\abovedisplayshortskip}{2pt plus 1pt}
\setlength{\belowdisplayshortskip}{3pt plus 1pt minus 1pt}

\section{Model and decay inputs}
\label{app:modelinputs}

\subsection{Spectroscopic inputs}
\label{app:spectroscopy}

The MGI Hamiltonian, smearing and spin interactions follow Refs.~\cite{Godfrey:1985xj,Song:2015fha,Song:2015nia}, with the inputs in Table~\ref{tab:parameters}. Alternative MGI and GI parameter sets are discussed in Refs.~\cite{Wang:2019mhs,Wang:2020prx,Liu:2024tgq,Liu:2026rja,Li:2023wgq,Sun:2026lup}. We retain the ${}^1L_J$--${}^3L_J$ spin-orbit mixing for $J=L$.

Screening describes unquenched effects phenomenologically, without explicit self-energy corrections from meson loops~\cite{Bai:2026atm}. The mass calculation supplies the individual MGI eigenfunctions used for parent and heavy daughter mesons in the decay amplitudes. Their radial and orbital structure is retained in the spatial overlaps, keeping the masses and widths in Table~\ref{tab:masswidth} consistent with the adopted spectroscopy inputs.

\begin{table}[!htb]
\caption{MGI parameters (GeV units where dimensional), with $\mu=0.03$ for $D$ and $0.02$ for $D_s$~\cite{Song:2015fha,Song:2015nia}.}
\label{tab:parameters}
\renewcommand{\arraystretch}{1.05}
\centering
\setlength{\tabcolsep}{0pt}
\begin{tabular*}{\linewidth}{@{\extracolsep{\fill}}lclc@{}}
\toprule
Parameter & Value & Parameter & Value\\
\midrule
$m_{u,d}$ & $0.22$ & $m_s$ & $0.419$\\
$m_c$ & $1.628$ & $b$ & $0.18~{\rm GeV}^2$\\
$c$ & $-0.253$ & $\mu$ & $0.03$\\
$\sigma_0$ & $1.80$ & $s_0$ & $1.55$\\
$\epsilon_c$ & $-0.168$ & $\epsilon_t$ & $0.025$\\
$\epsilon_{\rm so(v)}$ & $-0.035$ & $\epsilon_{\rm so(s)}$ & $0.055$\\
\bottomrule
\end{tabular*}
\end{table}

The remaining inputs are $(\alpha_1,\alpha_2,\alpha_3)=(0.25,0.15,0.20)$ and $(\gamma_1,\gamma_2,\gamma_3)=(1/2,\sqrt{10}/2,\sqrt{1000}/2)$~GeV. Spatial overlaps use numerical MGI wave functions for $D$ and $D_s$. Light mesons use simple harmonic oscillator (SHO) wave functions with $\beta_{\rm eff}=\sqrt{2\langle p^2\rangle_{1S}/3}$ from GI momentum moments at $\mu=0$~\cite{Godfrey:1985xj,Song:2015fha}; the nonstrange pseudoscalar and vector values are $0.77$ and $0.45$~GeV. Strong decays use theoretical heavy-meson masses and measured light-meson masses; the three-body fits use measured $D^0,D^-,\pi^+$ masses~\cite{ParticleDataGroup:2024cfk}.

\subsection{Strong decay widths}
\label{app:decays}

We use the QPC normalization of Refs.~\cite{Micu:1968mk,LeYaouanc:1972vsx,LeYaouanc:1973ldf,Song:2015fha}, with $\gamma=8.70$, charge multiplicities and a rate factor $1/3$ for $s\bar s$ creation. The partial waves give
\begin{equation}
 \Gamma(A\to BC)=\frac{\pi^2q}{M_A^2}\sum_{\ell,S}|\mathcal M^{\ell S}(q)|^2,
 \label{eq:qpcwidth}
\end{equation}
\noindent where $q$ is the two-body momentum and $\ell,S$ are the daughter orbital angular momentum and total spin. Distinct partial waves add incoherently; mixed-state component amplitudes are combined before squaring.

All charge-conserving, OZI-allowed $D_a(\pi,\eta,\eta',\rho,\omega)$ and $D_{s,a}K^{(*)}$ channels open at the theoretical parent mass are included, with MGI heavy daughters and all waves allowed by angular momentum and parity. Electromagnetic and multibody modes and broad-daughter convolutions are omitted.

For charge channels and partial waves labeled by $\alpha$, let $\Gamma^{(0)}_{R,c;\alpha}$ be the QPC width at $M_{R,c}$ and $q_{\alpha0}=q_\alpha(M_{R,c})$. Their sum is $\Gamma_{\rm sel}$; the inclusive $D\pi$ branching fraction is the sum of the two $D\pi$ charge-channel widths divided by $\Gamma_{\rm sel}$. We extrapolate using
\begin{equation}
\begin{aligned}
 \Gamma_{R,c}(m)&=\sum_{\alpha}\Gamma^{(0)}_{R,c;\alpha}
 \frac{M_{R,c}}{m}
 \left[\frac{q_\alpha(m)}{q_{\alpha0}}\right]^{2\ell_\alpha+1}\\
 &\quad\times\frac{\mathcal B_{\ell_\alpha}^{2}[q_\alpha(m)]}
 {\mathcal B_{\ell_\alpha}^{2}(q_{\alpha0})}
 \Theta(m-m_{\alpha1}-m_{\alpha2}),
\end{aligned}
\label{eq:widthenergy}
\end{equation}
\noindent where $m_{\alpha1,2}$ are daughter masses. Only phase space and barriers are continued: overlaps are not recalculated off shell, and this is not a complete unitary coupled-channel amplitude.

\section{Fitted coefficients}
\label{app:fitcoefficients}

Table~\ref{tab:coefficients} gives the coefficients for Schemes I--III. Here $\widetilde c_R$ is the $c_R$ of Eq.~\eqref{eq:amplitude} in the $\mathcal B_J(0)=1$ convention; the NR and $X(3790)$ coefficients are unaffected by the barrier rescaling. The overall amplitude scale and phase are conventional, and each spectrum has an independent normalization.

\begin{table}[!htb]
\caption{Coefficient moduli and phases $\phi=\arg\widetilde c$ (degrees), rounded to two decimal places. Unit $1$ denotes dimensionless coefficients.}
\label{tab:coefficients}
\centering\small
\renewcommand{\arraystretch}{1.0}
\setlength{\tabcolsep}{0pt}
\begin{tabular*}{\linewidth}{@{\extracolsep{\fill}}lrrc@{}}
\toprule
Contribution & $|\widetilde c|$ & $\phi$ (deg) & Unit\\
\midrule
\multicolumn{4}{c}{Scheme I}\\
$1\,{}^3S_1$ & $0.10$ & $-11.41$ & $\mathrm{GeV}^2$\\
$1\,{}^3P_2$ & $1.25$ & $2.25$ & $1$\\
$2\,{}^3S_1$ & $0.21$ & $93.90$ & $\mathrm{GeV}^2$\\
$1\,{}^3D_1$ & $3.97$ & $122.42$ & $\mathrm{GeV}^2$\\
$1\,{}^3D_3$ & $4.49$ & $-89.25$ & $\mathrm{GeV}^{-2}$\\
$3\,{}^3S_1$ & $11.49$ & $18.41$ & $\mathrm{GeV}^2$\\
NR & $0.92$ & $119.60$ & $1$\\
\midrule
\multicolumn{4}{c}{Scheme II}\\
$1\,{}^3S_1$ & $0.10$ & $31.33$ & $\mathrm{GeV}^2$\\
$1\,{}^3P_2$ & $1.25$ & $0.71$ & $1$\\
$2\,{}^3S_1$ & $0.29$ & $21.13$ & $\mathrm{GeV}^2$\\
$1\,{}^3D_1$ & $9.80$ & $-69.51$ & $\mathrm{GeV}^2$\\
$1\,{}^3D_3$ & $4.76$ & $46.76$ & $\mathrm{GeV}^{-2}$\\
$2\,{}^3D_1$ & $23.05$ & $172.26$ & $\mathrm{GeV}^2$\\
NR & $0.72$ & $-157.85$ & $1$\\
\midrule
\multicolumn{4}{c}{Scheme III}\\
$1\,{}^3S_1$ & $0.10$ & $80.44$ & $\mathrm{GeV}^2$\\
$1\,{}^3P_2$ & $1.26$ & $6.02$ & $1$\\
$2\,{}^3S_1$ & $0.44$ & $17.55$ & $\mathrm{GeV}^2$\\
$1\,{}^3D_1$ & $8.80$ & $-83.47$ & $\mathrm{GeV}^2$\\
$1\,{}^3D_3$ & $1.02$ & $40.97$ & $\mathrm{GeV}^{-2}$\\
$X(3790)$ & $8.00$ & $52.64$ & $\mathrm{GeV}^2$\\
NR & $0.61$ & $-98.93$ & $1$\\
\bottomrule
\end{tabular*}
\end{table}

\FloatBarrier
\clearpage
% Compact the APS reference-divider gaps; retain the native two-column handling.
\makeatletter
\def\bibsection{%
  \par
  \onecolumngrid@push
  \begingroup
    \baselineskip12\p@
    \bib@device{\textwidth}{245.5\p@}%
  \endgroup
  \nobreak\@nobreaktrue
  \addvspace{6\p@}%
  \par
  \onecolumngrid@pop
}
\makeatother
% Keep the bibliography font size; reduce only interline leading.
\makeatletter\g@addto@macro\bibfont{\setlength{\baselineskip}{9.0pt}}\makeatother
% Divide the current 33 references into balanced columns of 16 and 17 complete entries.
% Finish the previous entry's punctuation before starting reference 17.
\makeatletter
\@booleanfalse\balancelastpage@sw
\let\refs@bibitem@fin\bibitem@fin
\def\bibitem@fin{%
  \refs@bibitem@fin
  \ifnum\c@NAT@ctr=17\relax
    \par\newpage
    \setlength{\itemsep}{0pt}
  \fi
}
\makeatother
% Align the column bottoms with 0.64pt between left-column entries and 0pt on the right.
\setlength{\bibsep}{0.64pt}
\bibliography{x3790_refs.bib}

\begin{thebibliography}{33}%
\makeatletter
\providecommand \@ifxundefined [1]{%
 \@ifx{#1\undefined}
}%
\providecommand \@ifnum [1]{%
 \ifnum #1\expandafter \@firstoftwo
 \else \expandafter \@secondoftwo
 \fi
}%
\providecommand \@ifx [1]{%
 \ifx #1\expandafter \@firstoftwo
 \else \expandafter \@secondoftwo
 \fi
}%
\providecommand \natexlab [1]{#1}%
\providecommand \enquote  [1]{``#1''}%
\providecommand \bibnamefont  [1]{#1}%
\providecommand \bibfnamefont [1]{#1}%
\providecommand \citenamefont [1]{#1}%
\providecommand \href@noop [0]{\@secondoftwo}%
\providecommand \href [0]{\begingroup \@sanitize@url \@href}%
\providecommand \@href[1]{\@@startlink{#1}\@@href}%
\providecommand \@@href[1]{\endgroup#1\@@endlink}%
\providecommand \@sanitize@url [0]{\catcode `\\12\catcode `\$12\catcode
  `\&12\catcode `\#12\catcode `\^12\catcode `\_12\catcode `\%12\relax}%
\providecommand \@@startlink[1]{}%
\providecommand \@@endlink[0]{}%
\providecommand \url  [0]{\begingroup\@sanitize@url \@url }%
\providecommand \@url [1]{\endgroup\@href {#1}{\urlprefix }}%
\providecommand \urlprefix  [0]{URL }%
\providecommand \Eprint [0]{\href }%
\providecommand \doibase [0]{https://doi.org/}%
\providecommand \selectlanguage [0]{\@gobble}%
\providecommand \bibinfo  [0]{\@secondoftwo}%
\providecommand \bibfield  [0]{\@secondoftwo}%
\providecommand \translation [1]{[#1]}%
\providecommand \BibitemOpen [0]{}%
\providecommand \bibitemStop [0]{}%
\providecommand \bibitemNoStop [0]{.\EOS\space}%
\providecommand \EOS [0]{\spacefactor3000\relax}%
\providecommand \BibitemShut  [1]{\csname bibitem#1\endcsname}%
\let\auto@bib@innerbib\@empty
%</preamble>
\bibitem [{\citenamefont {Wang}\ \emph {et~al.}(2026)\citenamefont {Wang},
  \citenamefont {Liu},\ and\ \citenamefont {Gao}}]{Wang:2025dur}%
  \BibitemOpen
  \bibfield  {author} {\bibinfo {author} {\bibfnamefont {X.-F.}\ \bibnamefont
  {Wang}}, \bibinfo {author} {\bibfnamefont {X.}~\bibnamefont {Liu}},\ and\
  \bibinfo {author} {\bibfnamefont {Y.-N.}\ \bibnamefont {Gao}},\ }\bibfield
  {title} {\bibinfo {title} {{Colloquium: Hadron production in open-charm meson
  pairs at $e^+e^-$ colliders}},\ }\href {https://doi.org/10.1103/2mrp-chly}
  {\bibfield  {journal} {\bibinfo  {journal} {Rev. Mod. Phys.}\ }\textbf
  {\bibinfo {volume} {98}},\ \bibinfo {pages} {021001} (\bibinfo {year}
  {2026})}\BibitemShut {NoStop}%
\bibitem [{\citenamefont {Bai}\ \emph {et~al.}(2027)\citenamefont {Bai},
  \citenamefont {Chen}, \citenamefont {Huang}, \citenamefont {Liu},
  \citenamefont {Luo},\ and\ \citenamefont {Wang}}]{Bai:2026atm}%
  \BibitemOpen
  \bibfield  {author} {\bibinfo {author} {\bibfnamefont {Z.-Y.}\ \bibnamefont
  {Bai}}, \bibinfo {author} {\bibfnamefont {D.-Y.}\ \bibnamefont {Chen}},
  \bibinfo {author} {\bibfnamefont {Q.}~\bibnamefont {Huang}}, \bibinfo
  {author} {\bibfnamefont {X.}~\bibnamefont {Liu}}, \bibinfo {author}
  {\bibfnamefont {S.-Q.}\ \bibnamefont {Luo}},\ and\ \bibinfo {author}
  {\bibfnamefont {J.-Z.}\ \bibnamefont {Wang}},\ }\bibfield  {title} {\bibinfo
  {title} {{Unquenched charmonium and beyond}},\ }\href
  {https://doi.org/10.1016/j.physrep.2026.08.002} {\bibfield  {journal}
  {\bibinfo  {journal} {Phys. Rept.}\ }\textbf {\bibinfo {volume} {1204}},\
  \bibinfo {pages} {1} (\bibinfo {year} {2027})}\BibitemShut {NoStop}%
\bibitem [{\citenamefont {Ablikim}\ \emph {et~al.}(2026)\citenamefont {Ablikim}
  \emph {et~al.}}]{BESIII:2026emz}%
  \BibitemOpen
  \bibfield  {author} {\bibinfo {author} {\bibfnamefont {M.}~\bibnamefont
  {Ablikim}} \emph {et~al.} (\bibinfo {collaboration} {BESIII}),\ }\bibfield
  {title} {\bibinfo {title} {{Observation of a Charged Charmoniumlike Structure
  in $e^+e^-\to D^0D^-\pi^+ + \mathrm{c.c.}$}}\ }\href@noop {} {\  (\bibinfo
  {year} {2026})},\ \Eprint {https://arxiv.org/abs/2607.29138}
  {arXiv:2607.29138 [hep-ex]} \BibitemShut {NoStop}%
\bibitem [{\citenamefont {Flatt{\'e}}(1976)}]{Flatte:1976xu}%
  \BibitemOpen
  \bibfield  {author} {\bibinfo {author} {\bibfnamefont {S.~M.}\ \bibnamefont
  {Flatt{\'e}}},\ }\bibfield  {title} {\bibinfo {title} {{Coupled-channel
  analysis of the $\pi\eta$ and $K\bar{K}$ systems near $K\bar{K}$
  threshold}},\ }\href {https://doi.org/10.1016/0370-2693(76)90654-7}
  {\bibfield  {journal} {\bibinfo  {journal} {Phys. Lett. B}\ }\textbf
  {\bibinfo {volume} {63}},\ \bibinfo {pages} {224} (\bibinfo {year}
  {1976})}\BibitemShut {NoStop}%
\bibitem [{\citenamefont {Lu}\ \emph {et~al.}(2025)\citenamefont {Lu},
  \citenamefont {Song}, \citenamefont {Wang}, \citenamefont {Guo},
  \citenamefont {Li},\ and\ \citenamefont {Luo}}]{Lu:2025zae}%
  \BibitemOpen
  \bibfield  {author} {\bibinfo {author} {\bibfnamefont {J.-L.}\ \bibnamefont
  {Lu}}, \bibinfo {author} {\bibfnamefont {M.}~\bibnamefont {Song}}, \bibinfo
  {author} {\bibfnamefont {P.}~\bibnamefont {Wang}}, \bibinfo {author}
  {\bibfnamefont {J.-Y.}\ \bibnamefont {Guo}}, \bibinfo {author} {\bibfnamefont
  {G.}~\bibnamefont {Li}},\ and\ \bibinfo {author} {\bibfnamefont
  {X.}~\bibnamefont {Luo}},\ }\bibfield  {title} {\bibinfo {title} {{The bound
  and resonant states of $D^{(*)}D^{(*)}$ and $D^{(*)}{\bar{D}}^{(*)}$ with the
  complex scaling method}},\ }\href
  {https://doi.org/10.1140/epjc/s10052-025-14649-1} {\bibfield  {journal}
  {\bibinfo  {journal} {Eur. Phys. J. C}\ }\textbf {\bibinfo {volume} {85}},\
  \bibinfo {pages} {920} (\bibinfo {year} {2025})}\BibitemShut {NoStop}%
\bibitem [{\citenamefont {Ebert}\ \emph {et~al.}(2008)\citenamefont {Ebert},
  \citenamefont {Faustov},\ and\ \citenamefont {Galkin}}]{Ebert:2008kb}%
  \BibitemOpen
  \bibfield  {author} {\bibinfo {author} {\bibfnamefont {D.}~\bibnamefont
  {Ebert}}, \bibinfo {author} {\bibfnamefont {R.~N.}\ \bibnamefont {Faustov}},\
  and\ \bibinfo {author} {\bibfnamefont {V.~O.}\ \bibnamefont {Galkin}},\
  }\bibfield  {title} {\bibinfo {title} {{Excited heavy tetraquarks with hidden
  charm}},\ }\href {https://doi.org/10.1140/epjc/s10052-008-0754-8} {\bibfield
  {journal} {\bibinfo  {journal} {Eur. Phys. J. C}\ }\textbf {\bibinfo {volume}
  {58}},\ \bibinfo {pages} {399} (\bibinfo {year} {2008})}\BibitemShut
  {NoStop}%
\bibitem [{\citenamefont {Wang}(2024)}]{Wang:2024gvf}%
  \BibitemOpen
  \bibfield  {author} {\bibinfo {author} {\bibfnamefont {Z.-G.}\ \bibnamefont
  {Wang}},\ }\bibfield  {title} {\bibinfo {title} {{Ground states and first
  radial excitations of vector tetraquark states with explicit $P$-waves via
  QCD sum rules}},\ }\href {https://doi.org/10.1088/1674-1137/ad5ae5}
  {\bibfield  {journal} {\bibinfo  {journal} {Chin. Phys. C}\ }\textbf
  {\bibinfo {volume} {48}},\ \bibinfo {pages} {103103} (\bibinfo {year}
  {2024})}\BibitemShut {NoStop}%
\bibitem [{\citenamefont {Wang}(2021)}]{Wang:2021qus}%
  \BibitemOpen
  \bibfield  {author} {\bibinfo {author} {\bibfnamefont {Z.-G.}\ \bibnamefont
  {Wang}},\ }\bibfield  {title} {\bibinfo {title} {{Analysis of the vector
  hidden-charm tetraquark states without explicit $P$-waves via the QCD sum
  rules}},\ }\href {https://doi.org/10.1016/j.nuclphysb.2021.115592} {\bibfield
   {journal} {\bibinfo  {journal} {Nucl. Phys. B}\ }\textbf {\bibinfo {volume}
  {973}},\ \bibinfo {pages} {115592} (\bibinfo {year} {2021})}\BibitemShut
  {NoStop}%
\bibitem [{\citenamefont {Chen}\ \emph {et~al.}(2017)\citenamefont {Chen},
  \citenamefont {Chen}, \citenamefont {Liu}, \citenamefont {Liu},\ and\
  \citenamefont {Zhu}}]{Chen:2016spr}%
  \BibitemOpen
  \bibfield  {author} {\bibinfo {author} {\bibfnamefont {H.-X.}\ \bibnamefont
  {Chen}}, \bibinfo {author} {\bibfnamefont {W.}~\bibnamefont {Chen}}, \bibinfo
  {author} {\bibfnamefont {X.}~\bibnamefont {Liu}}, \bibinfo {author}
  {\bibfnamefont {Y.-R.}\ \bibnamefont {Liu}},\ and\ \bibinfo {author}
  {\bibfnamefont {S.-L.}\ \bibnamefont {Zhu}},\ }\bibfield  {title} {\bibinfo
  {title} {{A review of the open charm and open bottom systems}},\ }\href
  {https://doi.org/10.1088/1361-6633/aa6420} {\bibfield  {journal} {\bibinfo
  {journal} {Rept. Prog. Phys.}\ }\textbf {\bibinfo {volume} {80}},\ \bibinfo
  {pages} {076201} (\bibinfo {year} {2017})}\BibitemShut {NoStop}%
\bibitem [{\citenamefont {Navas}\ \emph {et~al.}(2024)\citenamefont {Navas}
  \emph {et~al.}}]{ParticleDataGroup:2024cfk}%
  \BibitemOpen
  \bibfield  {author} {\bibinfo {author} {\bibfnamefont {S.}~\bibnamefont
  {Navas}} \emph {et~al.} (\bibinfo {collaboration} {Particle Data Group}),\
  }\bibfield  {title} {\bibinfo {title} {{Review of particle physics}},\ }\href
  {https://doi.org/10.1103/PhysRevD.110.030001} {\bibfield  {journal} {\bibinfo
   {journal} {Phys. Rev. D}\ }\textbf {\bibinfo {volume} {110}},\ \bibinfo
  {pages} {030001} (\bibinfo {year} {2024})}\BibitemShut {NoStop}%
\bibitem [{\citenamefont {Song}\ \emph
  {et~al.}(2015{\natexlab{a}})\citenamefont {Song}, \citenamefont {Chen},
  \citenamefont {Liu},\ and\ \citenamefont {Matsuki}}]{Song:2015fha}%
  \BibitemOpen
  \bibfield  {author} {\bibinfo {author} {\bibfnamefont {Q.-T.}\ \bibnamefont
  {Song}}, \bibinfo {author} {\bibfnamefont {D.-Y.}\ \bibnamefont {Chen}},
  \bibinfo {author} {\bibfnamefont {X.}~\bibnamefont {Liu}},\ and\ \bibinfo
  {author} {\bibfnamefont {T.}~\bibnamefont {Matsuki}},\ }\bibfield  {title}
  {\bibinfo {title} {{Higher radial and orbital excitations in the charmed
  meson family}},\ }\href {https://doi.org/10.1103/PhysRevD.92.074011}
  {\bibfield  {journal} {\bibinfo  {journal} {Phys. Rev. D}\ }\textbf {\bibinfo
  {volume} {92}},\ \bibinfo {pages} {074011} (\bibinfo {year}
  {2015}{\natexlab{a}})}\BibitemShut {NoStop}%
\bibitem [{\citenamefont {Wang}\ \emph
  {et~al.}(2020{\natexlab{a}})\citenamefont {Wang}, \citenamefont {Chen},
  \citenamefont {Liu},\ and\ \citenamefont {Matsuki}}]{Wang:2020axi}%
  \BibitemOpen
  \bibfield  {author} {\bibinfo {author} {\bibfnamefont {J.-Z.}\ \bibnamefont
  {Wang}}, \bibinfo {author} {\bibfnamefont {D.-Y.}\ \bibnamefont {Chen}},
  \bibinfo {author} {\bibfnamefont {X.}~\bibnamefont {Liu}},\ and\ \bibinfo
  {author} {\bibfnamefont {T.}~\bibnamefont {Matsuki}},\ }\bibfield  {title}
  {\bibinfo {title} {{Universal non-resonant explanation to charmoniumlike
  structures $Z_c(3885)$ and $Z_c(4025)$}},\ }\href
  {https://doi.org/10.1140/epjc/s10052-020-08621-4} {\bibfield  {journal}
  {\bibinfo  {journal} {Eur. Phys. J. C}\ }\textbf {\bibinfo {volume} {80}},\
  \bibinfo {pages} {1040} (\bibinfo {year} {2020}{\natexlab{a}})}\BibitemShut
  {NoStop}%
\bibitem [{\citenamefont {Wang}\ \emph
  {et~al.}(2021{\natexlab{a}})\citenamefont {Wang}, \citenamefont {Chen},
  \citenamefont {Liu},\ and\ \citenamefont {Matsuki}}]{Wang:2020dmv}%
  \BibitemOpen
  \bibfield  {author} {\bibinfo {author} {\bibfnamefont {J.-Z.}\ \bibnamefont
  {Wang}}, \bibinfo {author} {\bibfnamefont {D.-Y.}\ \bibnamefont {Chen}},
  \bibinfo {author} {\bibfnamefont {X.}~\bibnamefont {Liu}},\ and\ \bibinfo
  {author} {\bibfnamefont {T.}~\bibnamefont {Matsuki}},\ }\bibfield  {title}
  {\bibinfo {title} {{Mapping a new cluster of charmoniumlike structures at
  $e^+ e^−$ collisions}},\ }\href
  {https://doi.org/10.1016/j.physletb.2021.136345} {\bibfield  {journal}
  {\bibinfo  {journal} {Phys. Lett. B}\ }\textbf {\bibinfo {volume} {817}},\
  \bibinfo {pages} {136345} (\bibinfo {year} {2021}{\natexlab{a}})}\BibitemShut
  {NoStop}%
\bibitem [{\citenamefont {Wang}\ \emph
  {et~al.}(2021{\natexlab{b}})\citenamefont {Wang}, \citenamefont {Zhou},
  \citenamefont {Liu},\ and\ \citenamefont {Matsuki}}]{Wang:2020kej}%
  \BibitemOpen
  \bibfield  {author} {\bibinfo {author} {\bibfnamefont {J.-Z.}\ \bibnamefont
  {Wang}}, \bibinfo {author} {\bibfnamefont {Q.-S.}\ \bibnamefont {Zhou}},
  \bibinfo {author} {\bibfnamefont {X.}~\bibnamefont {Liu}},\ and\ \bibinfo
  {author} {\bibfnamefont {T.}~\bibnamefont {Matsuki}},\ }\bibfield  {title}
  {\bibinfo {title} {{Toward charged $Z_{cs}(3985)$ structure under a
  reflection mechanism}},\ }\href
  {https://doi.org/10.1140/epjc/s10052-021-08877-4} {\bibfield  {journal}
  {\bibinfo  {journal} {Eur. Phys. J. C}\ }\textbf {\bibinfo {volume} {81}},\
  \bibinfo {pages} {51} (\bibinfo {year} {2021}{\natexlab{b}})}\BibitemShut
  {NoStop}%
\bibitem [{\citenamefont {Godfrey}\ and\ \citenamefont
  {Isgur}(1985)}]{Godfrey:1985xj}%
  \BibitemOpen
  \bibfield  {author} {\bibinfo {author} {\bibfnamefont {S.}~\bibnamefont
  {Godfrey}}\ and\ \bibinfo {author} {\bibfnamefont {N.}~\bibnamefont
  {Isgur}},\ }\bibfield  {title} {\bibinfo {title} {{Mesons in a Relativized
  Quark Model with Chromodynamics}},\ }\href
  {https://doi.org/10.1103/PhysRevD.32.189} {\bibfield  {journal} {\bibinfo
  {journal} {Phys. Rev. D}\ }\textbf {\bibinfo {volume} {32}},\ \bibinfo
  {pages} {189} (\bibinfo {year} {1985})}\BibitemShut {NoStop}%
\bibitem [{\citenamefont {Le~Yaouanc}\ \emph {et~al.}(1973)\citenamefont
  {Le~Yaouanc}, \citenamefont {Oliver}, \citenamefont {Pene},\ and\
  \citenamefont {Raynal}}]{LeYaouanc:1972vsx}%
  \BibitemOpen
  \bibfield  {author} {\bibinfo {author} {\bibfnamefont {A.}~\bibnamefont
  {Le~Yaouanc}}, \bibinfo {author} {\bibfnamefont {L.}~\bibnamefont {Oliver}},
  \bibinfo {author} {\bibfnamefont {O.}~\bibnamefont {Pene}},\ and\ \bibinfo
  {author} {\bibfnamefont {J.~C.}\ \bibnamefont {Raynal}},\ }\bibfield  {title}
  {\bibinfo {title} {{Naive quark pair creation model of strong interaction
  vertices}},\ }\href {https://doi.org/10.1103/PhysRevD.8.2223} {\bibfield
  {journal} {\bibinfo  {journal} {Phys. Rev. D}\ }\textbf {\bibinfo {volume}
  {8}},\ \bibinfo {pages} {2223} (\bibinfo {year} {1973})}\BibitemShut
  {NoStop}%
\bibitem [{\citenamefont {Le~Yaouanc}\ \emph {et~al.}(1974)\citenamefont
  {Le~Yaouanc}, \citenamefont {Oliver}, \citenamefont {Pene},\ and\
  \citenamefont {Raynal}}]{LeYaouanc:1973ldf}%
  \BibitemOpen
  \bibfield  {author} {\bibinfo {author} {\bibfnamefont {A.}~\bibnamefont
  {Le~Yaouanc}}, \bibinfo {author} {\bibfnamefont {L.}~\bibnamefont {Oliver}},
  \bibinfo {author} {\bibfnamefont {O.}~\bibnamefont {Pene}},\ and\ \bibinfo
  {author} {\bibfnamefont {J.~C.}\ \bibnamefont {Raynal}},\ }\bibfield  {title}
  {\bibinfo {title} {{Naive quark pair creation model and baryon decays}},\
  }\href {https://doi.org/10.1103/PhysRevD.9.1415} {\bibfield  {journal}
  {\bibinfo  {journal} {Phys. Rev. D}\ }\textbf {\bibinfo {volume} {9}},\
  \bibinfo {pages} {1415} (\bibinfo {year} {1974})}\BibitemShut {NoStop}%
\bibitem [{\citenamefont {Song}\ \emph
  {et~al.}(2015{\natexlab{b}})\citenamefont {Song}, \citenamefont {Chen},
  \citenamefont {Liu},\ and\ \citenamefont {Matsuki}}]{Song:2015nia}%
  \BibitemOpen
  \bibfield  {author} {\bibinfo {author} {\bibfnamefont {Q.-T.}\ \bibnamefont
  {Song}}, \bibinfo {author} {\bibfnamefont {D.-Y.}\ \bibnamefont {Chen}},
  \bibinfo {author} {\bibfnamefont {X.}~\bibnamefont {Liu}},\ and\ \bibinfo
  {author} {\bibfnamefont {T.}~\bibnamefont {Matsuki}},\ }\bibfield  {title}
  {\bibinfo {title} {{Charmed-strange mesons revisited: mass spectra and strong
  decays}},\ }\href {https://doi.org/10.1103/PhysRevD.91.054031} {\bibfield
  {journal} {\bibinfo  {journal} {Phys. Rev. D}\ }\textbf {\bibinfo {volume}
  {91}},\ \bibinfo {pages} {054031} (\bibinfo {year}
  {2015}{\natexlab{b}})}\BibitemShut {NoStop}%
\bibitem [{\citenamefont {Jacob}\ and\ \citenamefont
  {Wick}(1959)}]{Jacob:1959at}%
  \BibitemOpen
  \bibfield  {author} {\bibinfo {author} {\bibfnamefont {M.}~\bibnamefont
  {Jacob}}\ and\ \bibinfo {author} {\bibfnamefont {G.~C.}\ \bibnamefont
  {Wick}},\ }\bibfield  {title} {\bibinfo {title} {{On the General Theory of
  Collisions for Particles with Spin}},\ }\href
  {https://doi.org/10.1016/0003-4916(59)90051-X} {\bibfield  {journal}
  {\bibinfo  {journal} {Annals Phys.}\ }\textbf {\bibinfo {volume} {7}},\
  \bibinfo {pages} {404} (\bibinfo {year} {1959})}\BibitemShut {NoStop}%
\bibitem [{\citenamefont {Chung}(1998)}]{Chung:1997jn}%
  \BibitemOpen
  \bibfield  {author} {\bibinfo {author} {\bibfnamefont {S.~U.}\ \bibnamefont
  {Chung}},\ }\bibfield  {title} {\bibinfo {title} {{A General formulation of
  covariant helicity coupling amplitudes}},\ }\href
  {https://doi.org/10.1103/PhysRevD.57.431} {\bibfield  {journal} {\bibinfo
  {journal} {Phys. Rev. D}\ }\textbf {\bibinfo {volume} {57}},\ \bibinfo
  {pages} {431} (\bibinfo {year} {1998})}\BibitemShut {NoStop}%
\bibitem [{\citenamefont {Albaladejo}\ \emph {et~al.}(2020)\citenamefont
  {Albaladejo} \emph {et~al.}}]{JPAC:2020umo}%
  \BibitemOpen
  \bibfield  {author} {\bibinfo {author} {\bibfnamefont {M.}~\bibnamefont
  {Albaladejo}} \emph {et~al.} (\bibinfo {collaboration} {JPAC}),\ }\bibfield
  {title} {\bibinfo {title} {{$\omega \rightarrow 3\pi $ and $\omega \pi ^{0}$
  transition form factor revisited}},\ }\href
  {https://doi.org/10.1140/epjc/s10052-020-08576-6} {\bibfield  {journal}
  {\bibinfo  {journal} {Eur. Phys. J. C}\ }\textbf {\bibinfo {volume} {80}},\
  \bibinfo {pages} {1107} (\bibinfo {year} {2020})}\BibitemShut {NoStop}%
\bibitem [{\citenamefont {Von~Hippel}\ and\ \citenamefont
  {Quigg}(1972)}]{VonHippel:1972fg}%
  \BibitemOpen
  \bibfield  {author} {\bibinfo {author} {\bibfnamefont {F.}~\bibnamefont
  {Von~Hippel}}\ and\ \bibinfo {author} {\bibfnamefont {C.}~\bibnamefont
  {Quigg}},\ }\bibfield  {title} {\bibinfo {title} {{Centrifugal-barrier
  effects in resonance partial decay widths, shapes, and production
  amplitudes}},\ }\href {https://doi.org/10.1103/PhysRevD.5.624} {\bibfield
  {journal} {\bibinfo  {journal} {Phys. Rev. D}\ }\textbf {\bibinfo {volume}
  {5}},\ \bibinfo {pages} {624} (\bibinfo {year} {1972})}\BibitemShut {NoStop}%
\bibitem [{\citenamefont {Aaij}\ \emph {et~al.}(2016)\citenamefont {Aaij} \emph
  {et~al.}}]{LHCb:2016lxy}%
  \BibitemOpen
  \bibfield  {author} {\bibinfo {author} {\bibfnamefont {R.}~\bibnamefont
  {Aaij}} \emph {et~al.} (\bibinfo {collaboration} {LHCb}),\ }\bibfield
  {title} {\bibinfo {title} {{Amplitude analysis of $B^{-} \to D^{+} \pi^{-}
  \pi^{-}$ decays}},\ }\href {https://doi.org/10.1103/PhysRevD.94.072001}
  {\bibfield  {journal} {\bibinfo  {journal} {Phys. Rev. D}\ }\textbf {\bibinfo
  {volume} {94}},\ \bibinfo {pages} {072001} (\bibinfo {year}
  {2016})}\BibitemShut {NoStop}%
\bibitem [{\citenamefont {Ablikim}\ \emph {et~al.}(2020)\citenamefont {Ablikim}
  \emph {et~al.}}]{BESIII:2019dme}%
  \BibitemOpen
  \bibfield  {author} {\bibinfo {author} {\bibfnamefont {M.}~\bibnamefont
  {Ablikim}} \emph {et~al.} (\bibinfo {collaboration} {BESIII}),\ }\bibfield
  {title} {\bibinfo {title} {{Partial wave analysis of $\psi(3686)\to
  K^{+}K^{-}\eta$}},\ }\href {https://doi.org/10.1103/PhysRevD.101.032008}
  {\bibfield  {journal} {\bibinfo  {journal} {Phys. Rev. D}\ }\textbf {\bibinfo
  {volume} {101}},\ \bibinfo {pages} {032008} (\bibinfo {year}
  {2020})}\BibitemShut {NoStop}%
\bibitem [{\citenamefont {Nam}\ and\ \citenamefont {Ahn}(2026)}]{Nam:2026dmr}%
  \BibitemOpen
  \bibfield  {author} {\bibinfo {author} {\bibfnamefont {S.-i.}\ \bibnamefont
  {Nam}}\ and\ \bibinfo {author} {\bibfnamefont {J.~K.}\ \bibnamefont {Ahn}},\
  }\bibfield  {title} {\bibinfo {title} {{Three-body decay
  $\phi\to\pi^+\pi^-\pi^0$ with Omn{\`e}s-type final-state interactions}},\
  }\href {https://doi.org/10.1103/b56t-npm7} {\bibfield  {journal} {\bibinfo
  {journal} {Phys. Rev. D}\ }\textbf {\bibinfo {volume} {114}},\ \bibinfo
  {pages} {014022} (\bibinfo {year} {2026})}\BibitemShut {NoStop}%
\bibitem [{\citenamefont {Kim}\ \emph {et~al.}(2019)\citenamefont {Kim},
  \citenamefont {Park},\ and\ \citenamefont {Sahoo}}]{Kim:2018hlp}%
  \BibitemOpen
  \bibfield  {author} {\bibinfo {author} {\bibfnamefont {C.~S.}\ \bibnamefont
  {Kim}}, \bibinfo {author} {\bibfnamefont {S.~C.}\ \bibnamefont {Park}},\ and\
  \bibinfo {author} {\bibfnamefont {D.}~\bibnamefont {Sahoo}},\ }\bibfield
  {title} {\bibinfo {title} {{Angular distribution as an effective probe of new
  physics in semihadronic three-body meson decays}},\ }\href
  {https://doi.org/10.1103/PhysRevD.100.015005} {\bibfield  {journal} {\bibinfo
   {journal} {Phys. Rev. D}\ }\textbf {\bibinfo {volume} {100}},\ \bibinfo
  {pages} {015005} (\bibinfo {year} {2019})}\BibitemShut {NoStop}%
\bibitem [{\citenamefont {Wang}\ \emph {et~al.}(2019)\citenamefont {Wang},
  \citenamefont {Chen}, \citenamefont {Liu},\ and\ \citenamefont
  {Matsuki}}]{Wang:2019mhs}%
  \BibitemOpen
  \bibfield  {author} {\bibinfo {author} {\bibfnamefont {J.-Z.}\ \bibnamefont
  {Wang}}, \bibinfo {author} {\bibfnamefont {D.-Y.}\ \bibnamefont {Chen}},
  \bibinfo {author} {\bibfnamefont {X.}~\bibnamefont {Liu}},\ and\ \bibinfo
  {author} {\bibfnamefont {T.}~\bibnamefont {Matsuki}},\ }\bibfield  {title}
  {\bibinfo {title} {{Constructing $J/\psi$ family with updated data of
  charmoniumlike $Y$ states}},\ }\href
  {https://doi.org/10.1103/PhysRevD.99.114003} {\bibfield  {journal} {\bibinfo
  {journal} {Phys. Rev. D}\ }\textbf {\bibinfo {volume} {99}},\ \bibinfo
  {pages} {114003} (\bibinfo {year} {2019})}\BibitemShut {NoStop}%
\bibitem [{\citenamefont {Wang}\ \emph
  {et~al.}(2020{\natexlab{b}})\citenamefont {Wang}, \citenamefont {Qian},
  \citenamefont {Liu},\ and\ \citenamefont {Matsuki}}]{Wang:2020prx}%
  \BibitemOpen
  \bibfield  {author} {\bibinfo {author} {\bibfnamefont {J.-Z.}\ \bibnamefont
  {Wang}}, \bibinfo {author} {\bibfnamefont {R.-Q.}\ \bibnamefont {Qian}},
  \bibinfo {author} {\bibfnamefont {X.}~\bibnamefont {Liu}},\ and\ \bibinfo
  {author} {\bibfnamefont {T.}~\bibnamefont {Matsuki}},\ }\bibfield  {title}
  {\bibinfo {title} {{Are the $Y$ states around 4.6 GeV from $e^+e^-$
  annihilation higher charmonia?}},\ }\href
  {https://doi.org/10.1103/PhysRevD.101.034001} {\bibfield  {journal} {\bibinfo
   {journal} {Phys. Rev. D}\ }\textbf {\bibinfo {volume} {101}},\ \bibinfo
  {pages} {034001} (\bibinfo {year} {2020}{\natexlab{b}})}\BibitemShut
  {NoStop}%
\bibitem [{\citenamefont {Liu}\ \emph {et~al.}(2026{\natexlab{a}})\citenamefont
  {Liu}, \citenamefont {Man}, \citenamefont {Gao},\ and\ \citenamefont
  {Liu}}]{Liu:2024tgq}%
  \BibitemOpen
  \bibfield  {author} {\bibinfo {author} {\bibfnamefont {C.-X.}\ \bibnamefont
  {Liu}}, \bibinfo {author} {\bibfnamefont {Z.-L.}\ \bibnamefont {Man}},
  \bibinfo {author} {\bibfnamefont {T.-L.}\ \bibnamefont {Gao}},\ and\ \bibinfo
  {author} {\bibfnamefont {X.}~\bibnamefont {Liu}},\ }\bibfield  {title}
  {\bibinfo {title} {{Prospects for observing the missing $2D$ and $1F$
  charmonium states around 4~GeV}},\ }\href {https://doi.org/10.1103/lgfd-3pqc}
  {\bibfield  {journal} {\bibinfo  {journal} {Phys. Rev. D}\ }\textbf {\bibinfo
  {volume} {113}},\ \bibinfo {pages} {074009} (\bibinfo {year}
  {2026}{\natexlab{a}})}\BibitemShut {NoStop}%
\bibitem [{\citenamefont {Liu}\ \emph {et~al.}(2026{\natexlab{b}})\citenamefont
  {Liu}, \citenamefont {Wang},\ and\ \citenamefont {Liu}}]{Liu:2026rja}%
  \BibitemOpen
  \bibfield  {author} {\bibinfo {author} {\bibfnamefont {C.-X.}\ \bibnamefont
  {Liu}}, \bibinfo {author} {\bibfnamefont {J.-Z.}\ \bibnamefont {Wang}},\ and\
  \bibinfo {author} {\bibfnamefont {X.}~\bibnamefont {Liu}},\ }\bibfield
  {title} {\bibinfo {title} {{Exploring charmonia $\eta_c(6S,7S)$ in the
  $\Lambda_c^+\bar{\Lambda}_c^-$ invariant mass spectrum of $B^0\to
  K_S^0\Lambda_c^+\bar{\Lambda}_c^-$}},\ }\href
  {https://doi.org/10.1103/5lcd-6ml6} {\bibfield  {journal} {\bibinfo
  {journal} {Phys. Rev. D}\ }\textbf {\bibinfo {volume} {114}},\ \bibinfo
  {pages} {074050} (\bibinfo {year} {2026}{\natexlab{b}})}\BibitemShut
  {NoStop}%
\bibitem [{\citenamefont {Li}\ \emph {et~al.}(2023)\citenamefont {Li},
  \citenamefont {Li}, \citenamefont {Wang},\ and\ \citenamefont
  {Liu}}]{Li:2023wgq}%
  \BibitemOpen
  \bibfield  {author} {\bibinfo {author} {\bibfnamefont {X.-J.}\ \bibnamefont
  {Li}}, \bibinfo {author} {\bibfnamefont {Y.-S.}\ \bibnamefont {Li}}, \bibinfo
  {author} {\bibfnamefont {F.-L.}\ \bibnamefont {Wang}},\ and\ \bibinfo
  {author} {\bibfnamefont {X.}~\bibnamefont {Liu}},\ }\bibfield  {title}
  {\bibinfo {title} {{Spectroscopic survey of higher-lying states of $B_c$
  meson family}},\ }\href {https://doi.org/10.1140/epjc/s10052-023-12237-9}
  {\bibfield  {journal} {\bibinfo  {journal} {Eur. Phys. J. C}\ }\textbf
  {\bibinfo {volume} {83}},\ \bibinfo {pages} {1080} (\bibinfo {year}
  {2023})}\BibitemShut {NoStop}%
\bibitem [{\citenamefont {Sun}\ \emph {et~al.}(2026)\citenamefont {Sun},
  \citenamefont {Gao}, \citenamefont {Man},\ and\ \citenamefont
  {Liu}}]{Sun:2026lup}%
  \BibitemOpen
  \bibfield  {author} {\bibinfo {author} {\bibfnamefont {P.-Y.}\ \bibnamefont
  {Sun}}, \bibinfo {author} {\bibfnamefont {T.-L.}\ \bibnamefont {Gao}},
  \bibinfo {author} {\bibfnamefont {Z.-L.}\ \bibnamefont {Man}},\ and\ \bibinfo
  {author} {\bibfnamefont {X.}~\bibnamefont {Liu}},\ }\bibfield  {title}
  {\bibinfo {title} {{Proposed mixing between $2P$ and $1F$ wave charmonia}},\
  }\href {https://doi.org/10.1103/kpwy-d1ls} {\bibfield  {journal} {\bibinfo
  {journal} {Phys. Rev. D}\ }\textbf {\bibinfo {volume} {114}},\ \bibinfo
  {pages} {034011} (\bibinfo {year} {2026})}\BibitemShut {NoStop}%
\bibitem [{\citenamefont {Micu}(1969)}]{Micu:1968mk}%
  \BibitemOpen
  \bibfield  {author} {\bibinfo {author} {\bibfnamefont {L.}~\bibnamefont
  {Micu}},\ }\bibfield  {title} {\bibinfo {title} {{Decay rates of meson
  resonances in a quark model}},\ }\href
  {https://doi.org/10.1016/0550-3213(69)90039-X} {\bibfield  {journal}
  {\bibinfo  {journal} {Nucl. Phys. B}\ }\textbf {\bibinfo {volume} {10}},\
  \bibinfo {pages} {521} (\bibinfo {year} {1969})}\BibitemShut {NoStop}%
\end{thebibliography}%

\end{document}